# Electric field induced metallic behavior in thin crystals of ferroelectric $\alpha$-In$_2$Se$_3$

Justin R. Rodriguez[1,2], William Murray[3,2], Kazunori Fujisawa[1,2], Seng Huat Lee[1,2], Alexandra L. Kotrick[1,2], Yixuan Chen[1,2], Nathan Mckee[1,2], Sora Lee[4,2], Mauricio Terrones[1,2], Susan Trolier-McKinstry[4,2], Thomas N. Jackson[3,2], Zhiqiang Mao[1,2], Zhiwen Liu[3,2], Ying Liu[1,2],†

[1]*Department of Physics, Pennsylvania State University, University Park, PA 16802, USA.*

[2]*Materials Research Institute, Pennsylvania State University, University Park, PA 16802, USA.*

[3]*Department of Electrical Engineering, Pennsylvania State University, University Park, PA 16802, USA*

[4]*Department of Materials Science and Engineering, Pennsylvania State University, University Park, PA 16802, USA*

†Corresponding author. Email: yxl15@psu.edu

## Abstract

Ferroelectric semiconductor field effect transistors (FeSmFETs), which employ ferroelectric semiconducting thin crystals of $\alpha$-In$_2$Se$_3$ as the channel material as opposed to the gate dielectric in conventional ferroelectric FETs (FeFETs) were prepared and measured from room to the liquid-helium temperatures. These FeSmFETs were found to yield evidence for the reorientation of the electrical polarization and an electric field induced metallic state in $\alpha$-In$_2$Se$_3$. Our findings suggest that FeSmFETs can serve as a platform for the fundamental study of ferroelectric metals as well as the exploration of potential applications of semiconducting ferroelectrics.

Ferroelectricity is defined by the formation of spontaneous electrical polarization in a non-centrosymmetric crystal and the reorientation of the polarization between crystallographically defined directions by the application of an external electric field[1]. Ferroelectrics have been used to store data in either a capacitor or a ferroelectric field-effect transistor (FeFET)[2,3] configuration, the latter of which features a ferroelectric gate dielectric and a non-ferroelectric semiconducting channel. FeFETs provide not only fast and non-volatile data storage, but also a pathway towards "logic in memory" functions. However, the commercialization of FeFETs has encountered multiple obstacles ranging from retention time originating from the depolarization field and the gate leakage current[4] to endurance limited mainly by interface charge traps[5]. A FeFET variation which replaces the semiconducting channel in the conventional FET with a *ferroelectric* semiconducting channel but retains the non-ferroelectric gate dielectric was demonstrated recently[6]. This produces an "on" (or "off") state without an applied gate voltage, similar to a traditional FeFET. Such a device is referred to as a ferroelectric semiconductor field effect transistor (FeSmFET).

It is interesting to ask whether the "on" state in the FeSmFET can be a metallic state, which will make the transition from the "on" to the "off" state a ferroelectric metal-insulator transition. Anderson and Blount[7] examined the structural transition found in metallic $V_3Si$ and suggested that the formation of ionic displacements along a polar axis, which led to the loss of inversion symmetry, would be required for the occurrence of the apparent continuous phase transition seen in $V_3Si$. Furthermore, they suggested that such a state should be called a "ferroelectric metal." Being a metal appears to be inconsistent with the accepted definition of ferroelectricity as mobile charge carriers in a bulk metal will effectively screen any electric field, making the reorientation of the polarization unlikely. Nevertheless, metals showing the presence of electrical dipoles from ionic displacements along with a well-defined polar axis, such as $LiOsO_3$[8,9], $Ca_3Ru_2O_7$[10], and other materials[11], have attracted much attention in recent years even though the issue of whether the dipoles found in these materials are spontaneously ordered and reversable has not been resolved. Interestingly, the difficulty in reversing the polarization in a metal can be circumvented for a ferroelectric 2D crystal. A vertical electrical field, applied to a 2D crystal of 1T'-$WTe_2$ by a top and a bottom gate, was shown to lead to sharp jumps in sample conductance, which was attributed to polarization reversal[12]. However, direct evidence for ferroelectricity in 2D 1T'-$WTe_2$ is still lacking.

In$_2$Se$_3$, a layered transition metal chalcogenide (TMC) featuring a van der Waals interlayer coupling and an energy gap of 1.4 eV, was predicted[13] to be ferroelectric down to 1-unit-cell thickness in both $\alpha$- and $\beta$-phases. Supporting evidence for ferroelectricity in In$_2$Se$_3$ was found via piezoelectric force microscopy (PFM)[14,15,16,17,18] and second harmonic generation (SHG)[18,19,20], with a transition temperature up to 700 K[19]. Extensive device work carried out in the last few years also supports ferroelectricity in this layered TMC. An on/off state was found in rectifying devices by sweeping the source-drain voltage[14,15,16,17,21], revealing distinct hysteresis loops suggesting complex orientations of the polarization[13,14,16,19]. In the pioneering work of the FeSmFET featuring thin crystals of $\alpha$-In$_2$Se$_3$ and a gate dielectric of 90-nm-thick SiO$_2$ or 15-nm-thick HfO$_2$, respectively, large clockwise and counter-clockwise hysteresis loops were found in gate voltage sweeps[6]. The hysteresis was shown to persist down to 80 mK, making it unlikely that the observed hysteresis at such a low temperature is due to charge traps. These observations strongly support the existence of ferroelectricity in $\alpha$-In$_2$Se$_3$. However, the most explicit demonstration of ferroelectricity in $\alpha$-In$_2$Se$_3$ was obtained in a stacked capacitor/FET device[22]. This device consists of monolayer graphene on a SiO$_2$/Si substrate that functions as a bottom gate. The graphene is also the bottom electrode for a capacitor featuring a two-layer dielectric combining an insulating monolayer or bilayer of hexagonal boron nitride (hBN) and an atomically thin, semiconducting crystal of $\alpha$-In$_2$Se$_3$, which was covered by a top metal electrode/gate. The electric field applied between the graphene and the top electrode was used to reorient the polarization in $\alpha$-In$_2$Se$_3$. The graphene sandwiched between SiO$_2$ and hBN functioned as a charge detector through the position of the charge neutral point (CNP) in sample resistance *vs.* the gate voltage curves. From the clear and systematic shift in CNP as the polarization was flipped, a polarization value of 0.92 µC/cm$^2$ was estimated under an external field of 5 x 10$^5$ V/cm, which is reasonably close to the theoretically predicted[13] value of 0.6 µC/cm$^2$.

The in- and out-of-plane resistivities of $\alpha$-In$_2$Se$_3$ crystals grown by a modified Bridgman method used in this work, obtained from 4-point probe on bulk crystals using electrical contacts made by Ag paint, showed a variable-range hopping (VRH) conduction at low temperatures (below ~40 K, see Fig. 1a) after the unintentionally doped charge carriers are frozen out. Thin crystal of $\alpha$-In$_2$Se$_3$ were obtained by mechanical exfoliation from a bulk crystal and deposited onto a heavily doped silicon chip with 300-nm thick thermally grown surface of SiO$_2$. The thickness of a thin crystal of $\alpha$-In$_2$Se$_3$ was determined by atomic force microscope (AFM) after the transport measurements were carried out. Two types of FeSmFET featuring a Hall bar (Fig. 1e) and traditional FET (Fig. S1a in Supplementary Materials (SM)) pattern, respectively, were prepared by photolithography with electrodes of 5 nm of Ti and 45 nm of Au. The parameters for the four devices used in the present study are shown in Table S1 in SM. DC Electrical transport measurements were carried out in a Quantum Design Physical Property Measurement System (PPMS) equipped with a 9 T superconducting magnet that features a base temperature of 1.8 K. For temperature varying measurements, the device was cooled/warmed at zero gate voltage unless otherwise specified.

To characterize the thin crystals of $\alpha$-In$_2$Se$_3$, Raman spectroscopy, photoluminescence (PL) and second harmonic generation (SHG) measurements were used. The Raman spectra (Fig. 1b) confirmed that the crystals used were $\alpha$-In$_2$Se$_3$[23]. An energy gap value of 1.4 eV was revealed in the PL measurements (Fig. 1c), consistent with that found in the literature[23]. Strong SHG signals with the expected six-fold symmetry were also found (Fig. 1d), demonstrating that our $\alpha$-In$_2$Se$_3$ crystal indeed belongs to the *R3m* space group[17,18,19].

Source-drain current *vs*. voltage ($I_D$ *vs*. $V_{DS}$) characteristics were measured on the FeSmFETs at fixed gate voltages ($V_G$). The results for Sample A (Fig. 1e) with a channel length of 12 μm and thickness of 110 nm are shown in Fig. 2 for $V_G$ increasing from -75 to 75 V and back to -75V. No saturation in $I_D$ was observed in this range of the gate voltage up to 10 V for all $V_G$ values. Given that $I_D$ for negative $V_G$ is lower than $I_D$ for the positive, the $\alpha$-In$_2$Se$_3$ must be *n*-type, consistent with previous observations[24,25]. Similar features in $I_D$ *vs*. $V_{DS}$ characteristics were seen in other samples (Fig. S2). Interestingly, a marked change in the slope was found in most $I_D$ *vs*. $V_{DS}$ curves, showing that $I_D$ increases much faster at low $V_{DS}$ than at high $V_{DS}$ values. The sharp rise in $I_D$ at low $V_{DS}$ values (below a few tenth of volts) may be related to the the presence of two back-to-back Schottky diodes studied previously in other materials[26,27]. Behavior seen at high $V_{DS}$ values, in particular, in the linear plots (Fig. S2 and S3), is similar to what reported previously[6].

Clockwise transfer curves of $I_D$ *vs*. $V_G$, starting at $V_G$ = -75 V, were measured on our FeSmFET devices at fixed temperatures, *T*, from 300 to 2 K (Fig. 3 and Fig. S5). These results are consistent with those seen in the previous work[6]. The clockwise hysteresis loop points to the presence of a polarization in the *n*-type $\alpha$-In$_2$Se$_3$ semiconductor. Basically, at a sufficiently large negative $V_G$, say, -75 V, the downward pointing electric field will force the polarization inside the $\alpha$-In$_2$Se$_3$ crystal downward (Figs. S1), resulting in *positive* bound surface charge on the *bottom* surface of the crystal due to the presence of the polarization. Consequently, the energy bands will band upward (Fig. S1). The gate voltage induced *positive* charge carriers will deplete the conduction band (the existing negative charge carriers will be "drained" from the channel), which will shut down conduction channels between the source and the drain. On the *top* of the $\alpha$-In$_2$Se$_3$ crystal, however, the *negative* bound charge from the downward pointing polarization will push down the conduction band, placing the Fermi energy within the conduction band (Fig. S1). However, even though the low density of the gate voltage-induced *positive* charge carriers they cannot deplete the conduction band fully because the gate electric field is weak on the top surface of the crystal, no conduction channel between the source and the drain is expected there either because of the low carrier density. An "off" state of the FeSmFET is thus expected, which was indeed observed.

As $V_G$ increased from the -75 V to 0 and then 75 V, the polarization will start to reverse locally. The depletion layer on the bottom surface of the $\alpha$-In$_2$Se$_3$ crystal will be reduced, helping push down the conduction band. A conduction channel between the source and the drain will eventually be established, leading to the "on" state of the FeSmFET. The device will continue to be in the "on" state as $V_G$ is increased further to 75 V. Now the polarization will switch to point upward so that the bound charge from the polarization will be negative on the bottom surface of the crystal, featuring negative mobile charge carriers induced by the positive gate voltage. Ramping $V_G$ from 75 V to 0, the polarization will turn downwards locally, leading to *positive* bound charge from the polarization on the *bottom* surface of the crystal. The existing and gate induced *negative* charge carriers could be bound to the *positive* surface charge from the polarization, creating local areas that are non-conducting. As the gate voltage decreases further, the polarization will continue to flip, leading to continued growth of non-conducting areas and a decreasing $I_D$. Eventually all conduction channels disappear, leading to vanishingly small $I_D$. A clockwise hysteresis loop as shown in Fig. 3 will be obtained. Our observation is therefore fully consistent with the existence of polarization in $\alpha$-In$_2$Se$_3$, as argued previously[6].

The overall hysteresis decreased as the temperature *T* was lowered (Fig. 3). Thus, it is natural to ask whether the reduction in hysteresis was a result of a change in coercive field as the temperature was lowered. This seems to be unlikely given that the ferroelectric transition temperature, T$_c$, of $\alpha$-In$_2$Se$_3$ was reported to be 700 K[19] as the coercive field of a ferroelectric material would increase as *T* is lowered below T$_c$[28,29] or stay as a constant far below it. The more likely scenario is the decrease in hysteresis as *T* was lowered was due to the presence of charge traps in our sample.

Charge traps in FETs are known to lead to a hysteresis loop. At higher temperatures, the hysteresis originating from charge traps and that from polarization appear to coexist in our samples. However, at a liquid-helium temperature, at which the binding and unbinding of mobile charge carriers from their traps are expected to be suppressed, the observed hysteresis should be only due to the reversal of the polarization, as argued previously[6].

Values of two-point resistance $R_{DS}$ (= $I_D/V_{DS}$, taken from the top of the hysteresis loop) are plotted against $T$ in Fig. 4a, showing decreasing $R_{DS}$ with the lowering $T$ and the emergence of a metallic state. The four-point sample resistance, $R(T)$, of the same crystal was also measured as a function of $T$ (inset of Fig. 4a), showing that $R_{DS}(T)$ and $R(T)$ have similar behavior. This suggests that the contact resistance between Ti/Au and $\alpha$-In$_2$Se$_3$, which was measured at the room temperature (Fig. S7), did not make a big difference in the behavior of $I_D$. Data obtained for $V_G$ = 75 V in samples B and C showed a positive $dR/dT$ at higher temperatures and a complete flattening-off in $R_{DS}(T)$ down to 4 K (Figs. 4b and S6). A small negative $dR/dT$ was seen at lowest temperatures in Sample A, even at $V_G$ = 75 V where the density of gate voltage induced mobile charge carriers is the largest, appears to be due to sample specific disorder. Weak localization in a weakly disordered metallic sample can lead to a negative $dR/dT$ when $T$ is sufficiently low[30]. The maximum 2D electric conductivity obtained from the four-point measurements was found to be around 80 $\sigma_0$, where $\sigma_0$ = e$^2$/h (= 4.08 x 10$^{-5}$Ω$^{-1}$) is the quantum conductance, e is the electron charge and h is Planck's constant. Above $\sigma_0$, a negative $dR/dT$ is expected due to weak localization, along with positive or negative magnetoconductance (MC) depending on the strength of the spin-orbital coupling[30]. Our measurements showed positive MC at 1.8 K (Fig. 4c), as well as 10 K and 50 K (data not shown), consistent with the weak spin-orbital coupling expected for $\alpha$-In$_2$Se$_3$. The MC data was shown in Fig. 4c to fit Maekawa-Fukuyama theory of 2D weak localization[31] quantitatively.

In the metallic state, the negative mobile charge carriers should be accumulated near the bottom of the $\alpha$-In$_2$Se$_3$ crystal while the rest of the crystal remains semiconducting. This layer of mobile charge will tend to screen the gate electric field, making the polarization in the semiconducting region of the crystal less affected by the gate electric field. However, as shown in the data, some of polarization can still be reoriented by the field even in the metallic state. As a result, the 2D metallic sheet of electrons and the polarization must coexist in $\alpha$-In$_2$Se$_3$.

The interesting question is how the bound charge of polarization on the crystal bottom will affect the accumulation of the mobile electrons. At $V_G = 75$ V the electric field is expected to induce an electron density of 5 x $10^{12}$/cm$^2$. Hall measurements showed that the Hall voltage is a linear function of the magnetic field, suggesting that only electrons are present in the sample. The density of electrons at 4 K is roughly what would be expected solely from that induced by the gate electric field, unaffected by the polarization (Fig. 4d). At higher temperatures however, the electron density obtained by the Hall measurements is larger than that induced by the gate voltage. This is reasonable, as the existing unintentionally doped electrons that are bound to the positive charge traps at low temperatures would start to be released as the temperature was raised, consistent with the observation of the broadening hysteresis noted above.

The observations presented above demonstrate a well-functioning FET with a large $I_D$ even when gate voltage is at zero. Such a FeSmFET can be used for logic operations as well as a memory device in the microelectronic circuitry with the "logic in memory" functionalities. In addition, ultrathin $\alpha$-In$_2$Se$_3$ was shown to provide a testbed for fundamental research on ferroelectric metals as well as ferroelectric metal-insulator transitions tuned by a gate voltage.

**Acknowledgement**. Work is supported by the NSF (Grant No. EFMA1433378). The $\alpha$-In$_2$Se$_3$ crystals used in this study were produced by the Penn State 2D Crystal Consortium – Materials Innovation Platform (2DCC-MIP) under NSF cooperative agreement DMR-1539916.

**Data Availability.** The data that support the findings of this study are available from the corresponding author upon reasonable request.

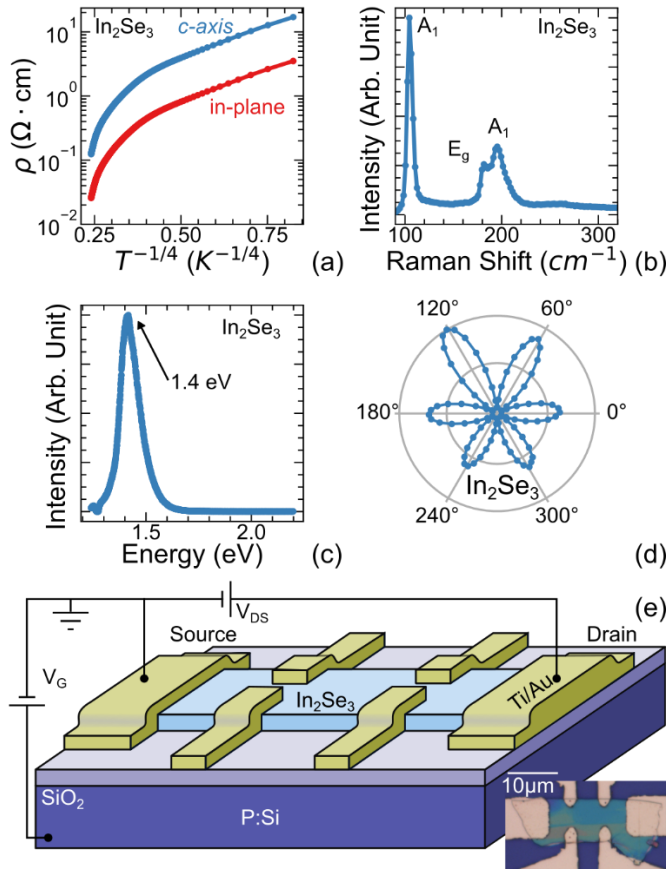

**Fig. 1**. (a) In- and out-of-plane resistivities of bulk single-crystals α-In$_2$Se$_3$ showing Mott variable-range hopping conduction behavior, ρ$_{\text{in-plane, c-axis}}$ (T) ~ exp[(T$_0$/T)$^{1/4}$], where T$_0$ is a constant, below around T = 160 K. Also shown are results of Raman spectroscopy (b), photoluminescence (c), and second harmonic generation (d) measurements. A schematic of an FeSmFET in the Hall bar pattern is shown in (e). Inset: Optical image of a FeSmFET device following this design with the 10-µm scale bar also shown.

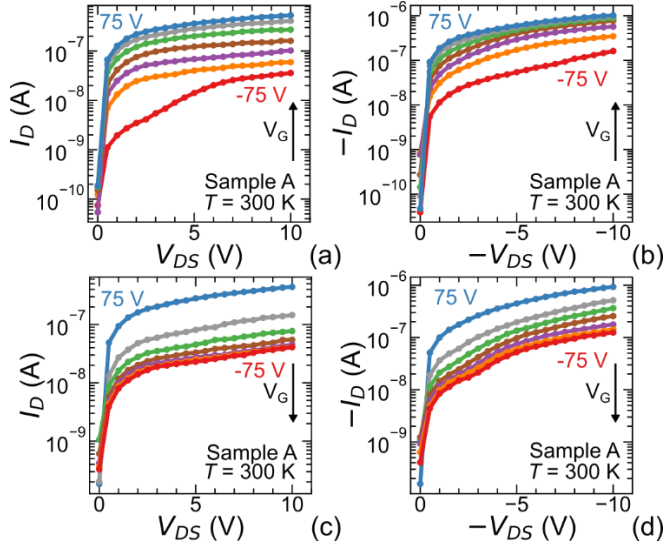

**Fig. 2**. Curves of $I_D$-$V_{DS}$ for Sample A obtained at room temperature for fixed gate voltages in an increasing order, $V_G$ = -75, -50, -25, 0, 25, 50, 75 V at a positive (a) and negative (b) values of $V_{DS}$. Corresponding $I_D$-$V_{DS}$ curves for decreasing $V_G$ at the same gate voltages are shown in (c) and (d).

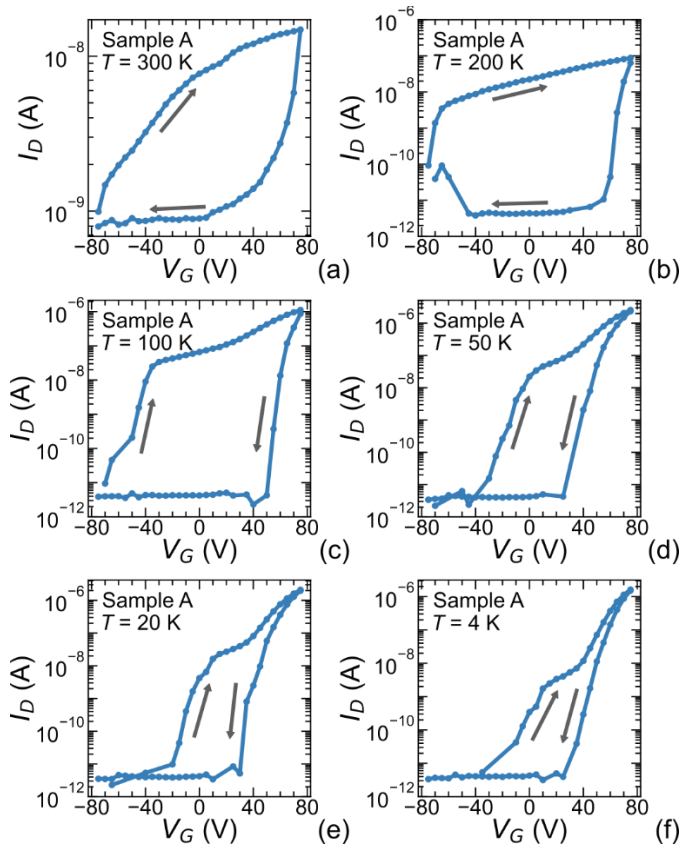

**Fig. 3**. (a-f) Transfer curves of source-drain current *vs.* gate voltage, $I_D$ *vs.* $V_G$, obtained at fixed temperatures (*T*) as indicated. $V_G$ was initially decreased from 0 to -75 V, after which $V_G$ was ramped from –75 to 75 V and then back to –75 V while the $I_D$ vs. $V_G$ curve was measured. A clockwise hysteresis loop was obtained at each temperature.

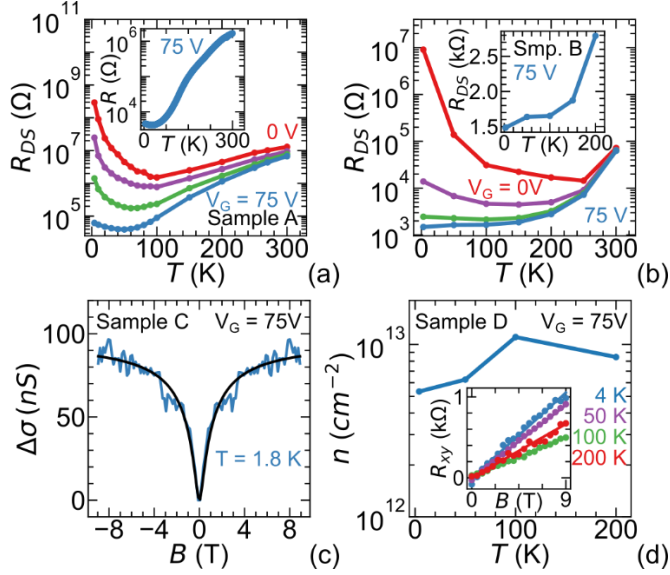

**Fig. 4**. (a) Two-point sample resistance, $R_{SD}(T)$ ($R_{DS} = I_D/V_{DS}$) obtained at $V_G = 0$, 25, 50, and 75 V, in the top portion of the clockwise hysteresis loops taken at fixed temperatures for Sample A. Inset: Four-point resistance vs. temperature, $R(T)$, measured while the device was warming up after it was cooled to 4 K at $V_G = -75$ V followed by ramping $V_G$ from -75 to 75 V at 4 K. (b) $R_{DS}(T)$ for Sample B in semi-log plot obtained from the top portion of clockwise hysteresis loops. Inset: $R_{DS}(T)$ for $V_G = 75$ V in linear plot. (c) Magnetoconductance at $T = 1.8$ K and $V_G = 75$ V for Sample C. The Maekawa-Fukuyama theory was shown to fit the data (see main text). (d) Charge carrier density, $n$, as a function of $T$ for Sample D. The device design for Samples A and D is shown in Fig. 1e and that for Samples B and C is shown in Fig. S1 (a).

**Supplementary Materials**

**Table S1**. Parameters for 4 devices studied in the current work, where t is the thickness of the α-In$_2$Se$_3$ crystal, L is the channel length (length between the voltage leads), w is the channel width, and DVG is the size of the hysteresis loop measured at $I_D$ = 1 nA and $T$ = 300 K.

| Sample | t (nm) | L (μm) | w (μm) | Device pattern | $\Delta V_G$ (V) at 1nA and 300K |
|---|---|---|---|---|---|
| A | 110 | 14 | 10 | Hall bar | 87V |
| B | 13 | 5 | 30 | Standard FET | 19V |
| C | 20 | 2 | 5 | Standard FET | 23V |
| D | 110 | 14 | 9 | Hall bar | not measured |

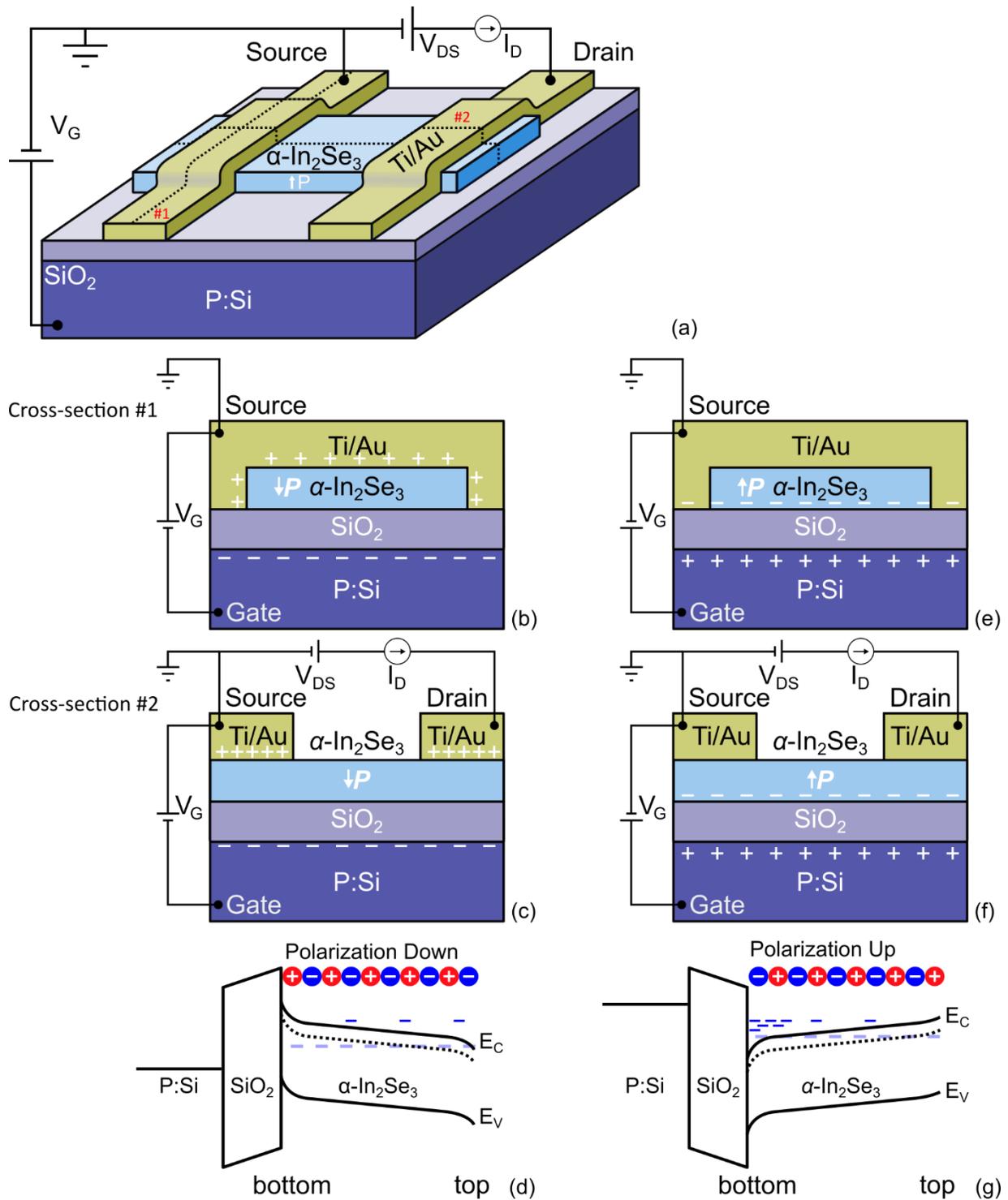

**Fig. S1.** a) Schematic of FeSmFET adopted in Samples B and C. b-c) Cross sectional views of the device cut along the dashed lines marked by #1 and #2 in a), perpendicular to and along the direction of $I_D$, respectively, for $V_{SD} = 0.1$ V and $V_{SD} = -75$. The polarization is pointing downwards.

d) Band diagrams along the vertical direction in (c) near the center of the $a$-In$_2$Se$_3$ channel is shown for $V_{SD}$ = -75 V. The mobile charge carriers are marked by +/- signs while the bound charge from electrical polarization is marked by the +/- circles. The crystal of $\alpha$-In$_2$Se$_3$ is unintentionally doped and $n$-type with the impurity states formed slightly below the bottom of the conduction band (black dashed line). The band bending on the top and bottom surfaces of the $a$-In$_2$Se$_3$ crystal is assumed to be induced by only the bound surface charge from the polarization. The location of the Fermi level (dashed blue line) is determined by the existing charge carriers. The bottom layer of the crystal is depleted at $V_{SD}$ = -75 V. e-f) Cross sectional views of the device cut along the dashed lines marked by #1 and #2 in a), respectively for $V_{SD}$ = 0.1 V and $V_{SD}$ = 75. The polarization is pointing upwards. g) The band along the vertical direction in (f) near the center of the $a$-In$_2$Se$_3$ channel is shown for $V_{SD}$ = 75 V. The location of the Fermi level is indicated by blue dashed line. The impurity states are formed slightly below the bottom of the conduction band (black dashed line). The accumulation of mobile electrons near the bottom surface of the crystal, the tilting of the energy band due to the existing of a finite electric field, and the existence of electrons from the impurity states in the interior of the $a$-In$_2$Se$_3$ crystal are shown schematically, when appropriate.

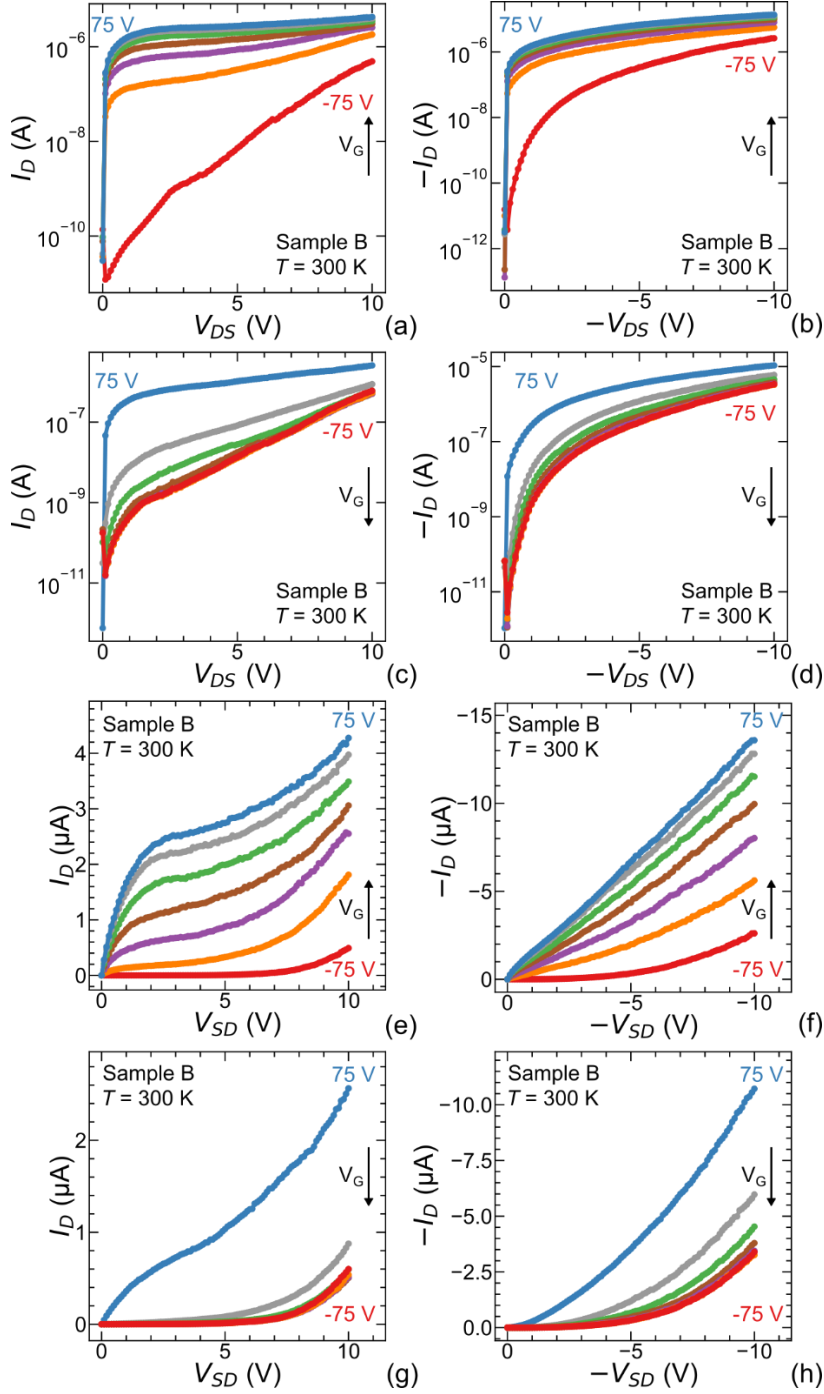

**Fig. S2.** $I_D$ *vs.* $V_{DS}$ measured in Sample B. $V_G$ was fixed for each curve in 25 V increments from -75 V to 75 V, the same as that shown in Fig. 2 in the main text, for positive (a) and negative (b) $V_{DS}$. c-d). Corresponding positive and negative $I_D$ *vs.* $V_{DS}$ while $V_G$ is decreased from 75 V to -75 V in 25 V increments. All measurements were done at 300 K. The linear plots of the same data are shown in e-h).

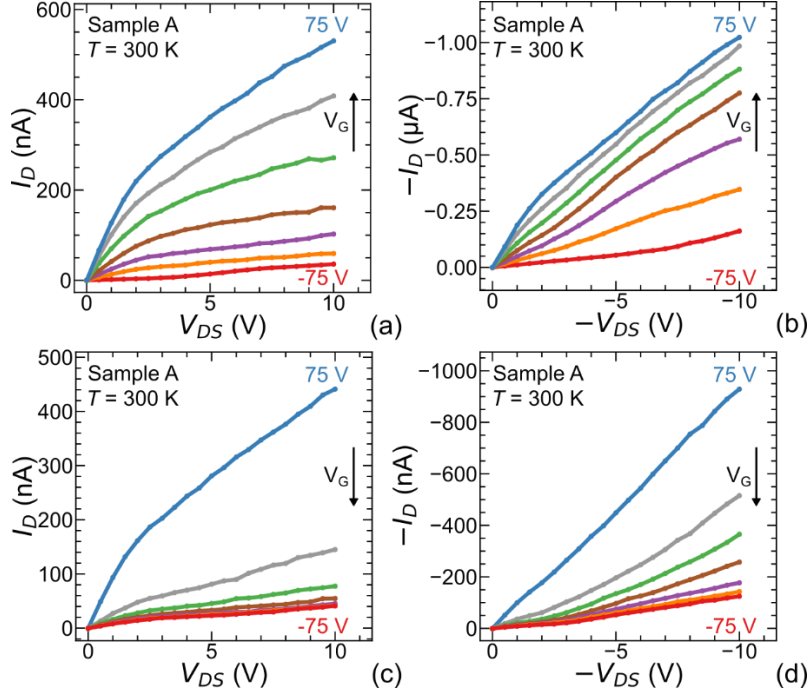

**Fig. S3.** $I_D$ vs. $V_{DS}$ data obtained in Sample A (shown in Fig. 2 of the main text in a semilog plot) plotted in the linear scale. Gate voltages $V_G$ was in an increasing order, $V_G$ = -75, -50, -25, 0, 25, 50, 75 V at a positive (a) and negative (b) values of $V_{DS}$. Corresponding $I_D$-$V_{DS}$ curves for decreasing $V_G$ at the same gate voltages are shown in (c) and (d). All measurements were done at 300 K.

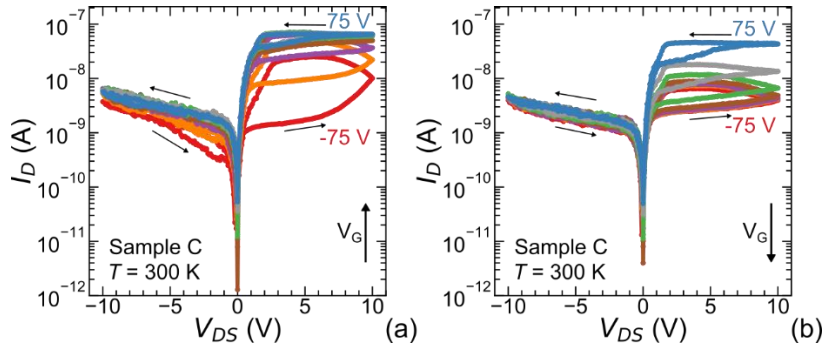

**Fig. S4.** $I_D$ *vs*. $V_{DS}$ measured in Sample C with increasing and decreasing gate voltage measured at $T = 300$ K. The sign of the $I_D$ reversed for negative $V_{DS}$ to facilitate the semi-log plotting. Gate voltage $V_G$ was increased in (a) and decreased in (b) in 25 V increments from -75 V to 75 V. After $V_G$ was set, the $V_{DS}$ was increased then decreased from 0 to -10V and back (not shown for clarity), then from 0 V, to 10 V, to -10 V and finally to 0 V. Note that the maximum current is smaller than shown in Fig. S6 below as the data was taken after the sample was left at room temperature for an extended time, resulting in a small degradation of the sample from previous measurements.

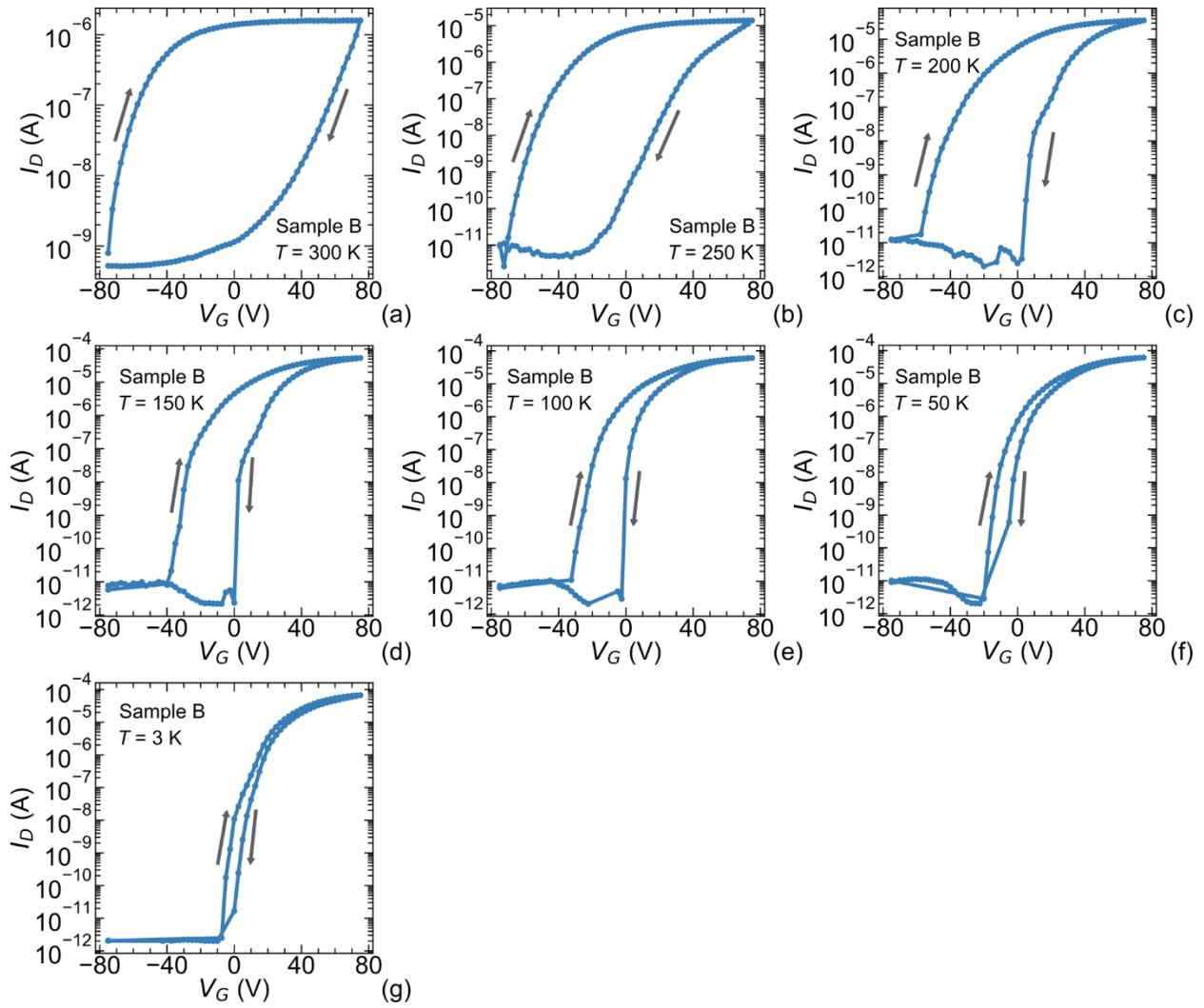

**Fig. S5.** a-g) The transfer curves of $I_D$ vs. $V_G$ of Sample B, performed using the same procedure as Sample A (shown in Fig. 3 in the main text). $V_{DS} = 0.1$ V for all measurements. $V_G = 25$ V, 50 V, and 75 V for upper portion of loop is used for Fig. 4b in the main text.

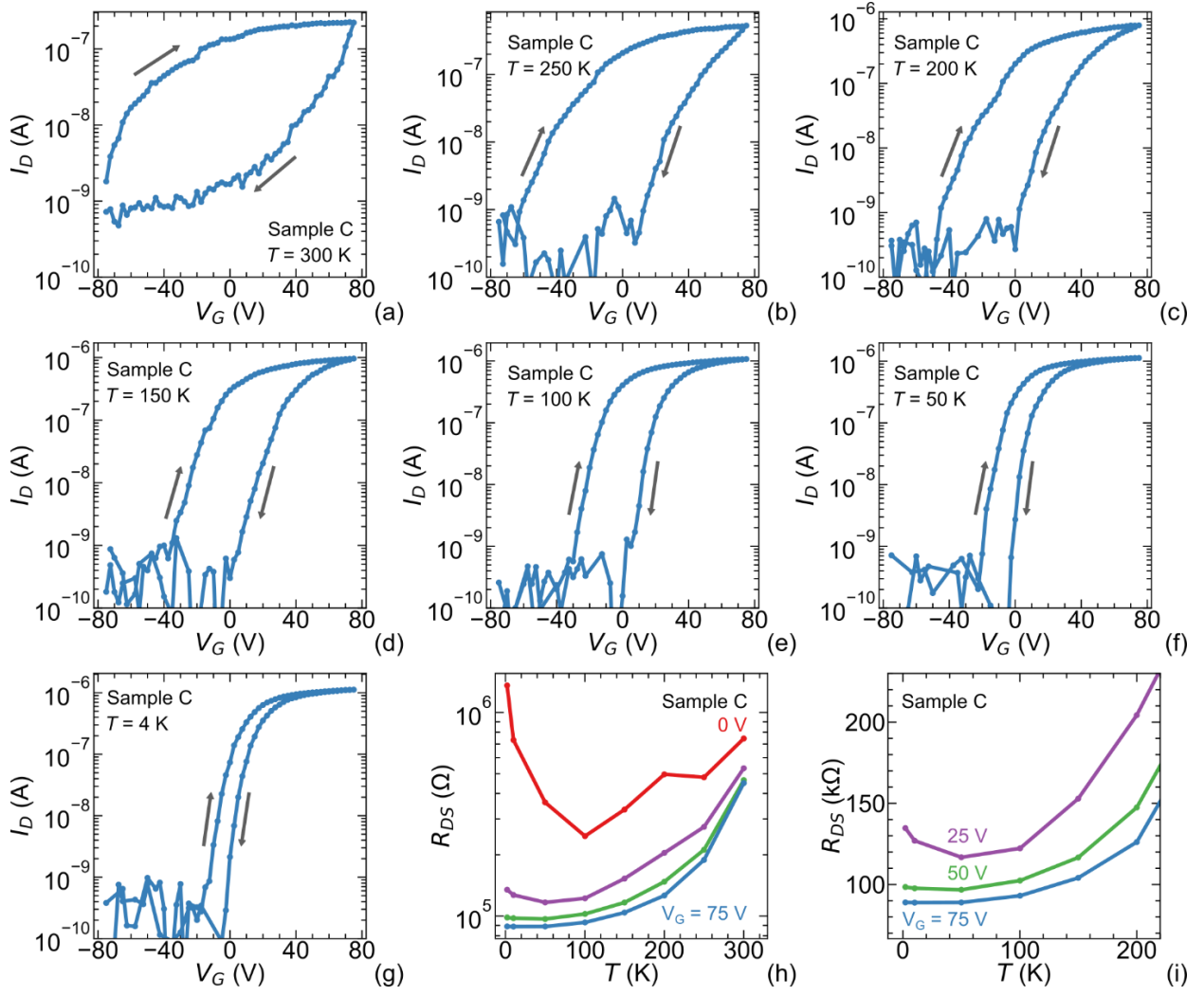

**Fig. S6**. a-g) The transfer curves of $I_D$ *vs.* $V_G$ of Sample C performed using the same procedure as Sample A shown in Fig. 3 in the main text; $V_{DS} = 0.1$ V for all measurements. h). $R_{DS}(T, V_G)$ taken from the clockwise hysteresis loops of $I_D$ *vs.* $V_G$ curves a-g, showing the same behavior seen in Figs. 4a-b in the main text. i) Low-temperature data from (h) shown in a linear plot for comparison.

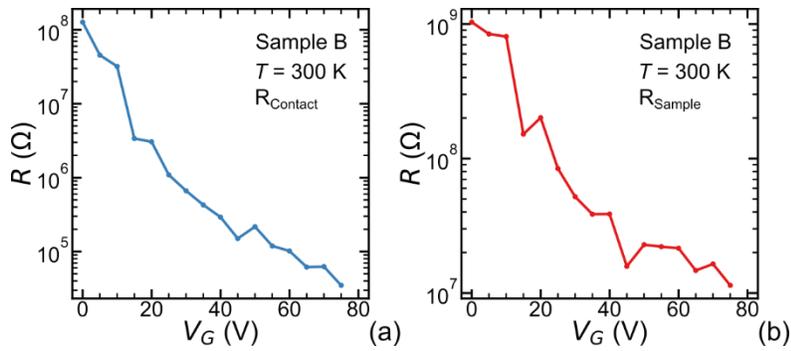

**Fig. S7.** a) Contact resistance between α-In$_2$Se$_3$ and Ti/Au leads measured with a $V_{DS}$ = 0.1 V bias between source and drain leads using a 3-point lead configuration at room temperature. b) 4-point resistance of the sample.

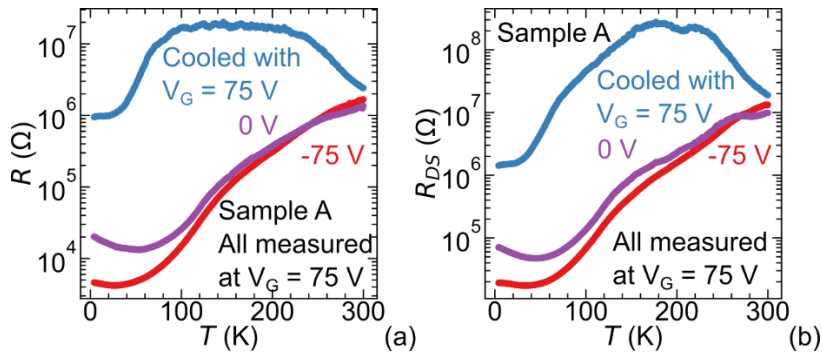

**Fig. S8**. History dependence of $R$(T) and $R_{DS}$(T) of Sample A. $V_G$ was ramped from 0 to – 75 V at 300 K, then cooled to 4K with $V_G$ ramped to and fixed at -75 V, 0 V, or 75V. At $T$ = 4 K, $V_G$ was ramped again going through the full hysteresis loop clockwise to $V_G$ = 75 V before $R$ was measured as a function of increasing $T$ while $V_G$ is fixed at that value. $R$ is seen to the smallest when the sample was cooled at $V_{G\,=}$ -75 V.

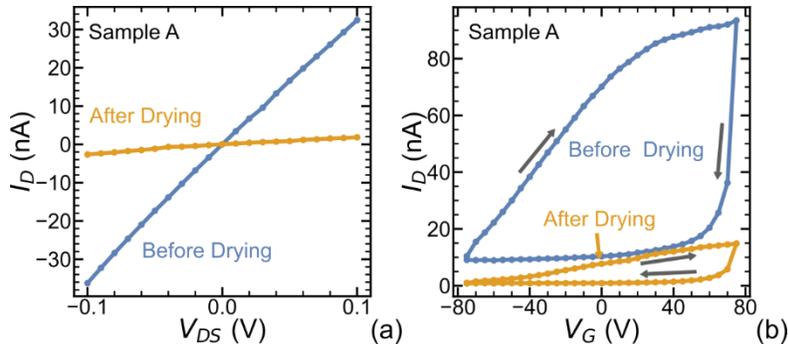

**Fig. S9.** Detection of charge traps at the interface between α-$In_2Se_3$ and $SiO_2$. a) $I_D$-$V_{DS}$, curves for Sample A before and after the "drying" - The sample was "dried" by heating to 400 K in high vacuum for one hour within the PPMS was kept at 400 K for an hour to remove water molecules trapped at the interface. b) The transfer curves of source-drain current *vs.* gate voltage, $I_D$ *vs.* $V_G$, at 300 K before and after the "drying". The measurements were taken immediately before and after "drying".

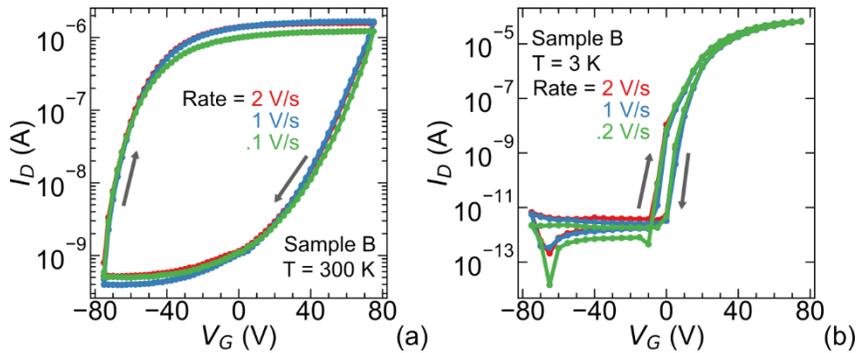

**Fig. S10.** Rate dependence on hysteresis loops of Sample B. Transfer curves of $I_D$ *vs.* $V_G$ take at different sweeping rates at 300 K (a) and 3 K (b).

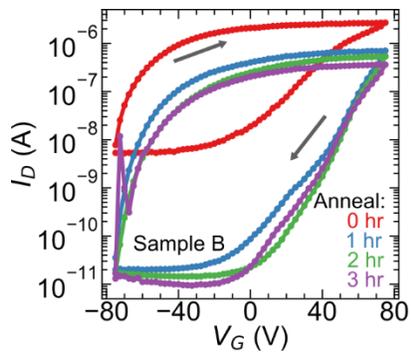

**Fig. S11.** Transfer curves of source-drain current *vs*. gate voltage, $I_D$ *vs*. $V_G$, at 300 K before and after the sample was dried by heating to 400 K in a high vacuum for one hour in the PPMS then cooled down to room temperature to be remeasured. This was repeated two additional times for a total of 3 hours of heating at 400 K.

[20] Yu Zhou, Di Wu, Yihan Zhu, Yujin Cho, Qing He, Xiao Yang, Kevin Herrera, Zhaodong Chu, Yu Han, Michael C. Downer, Hailin Peng, and Keji Lai. Out-of-Plane Piezoelectricity and Ferroelectricity in Layered α-In$_2$Se$_3$ Nanoflakes. *Nano Lett.* 2017 **17**(9), 5508–5513 (2017). https://doi.org/10.1021/acs.nanolett.7b02198.

[21] Huai Yang, Mengqi Xiao, Yu Cui, Longfei Pan, Kai Zhao, and Zhongming Wei. Nonvolatile memristor based on heterostructure of 2D room-temperature ferroelectric α-In$_2$Se$_3$ and WSe$_2$. *Sci. China Inf. Sci.* 2019 **62**(12), 1–8 (2019). https://doi.org/10.1007/s11432-019-1474-3.

[22] Siyuan Wan, Yue Li, Wei Li, Xiaoyu Mao, Chen Wang, Chen Chen, Jiyu Dong, Anmin Nie, Jianyong Xiang, Zhongyuan Liu, Wenguang Zhu, and Hualing Zeng. Nonvolatile Ferroelectric Memory Effect in Ultrathin α-In$_2$Se$_3$. *Advanced Functional Materials* 2019 **29**(20), 1808606 (2019). https://doi.org/10.1002/adfm.201808606.

[23] Nilanthy Balakrishnan, Elisabeth D. Steer, Emily F. Smith, Zakhar R. Kudrynskyi, Zakhar D. Kovalyuk, Laurence Eaves, Amalia Patanè, and Peter H. Beton. Epitaxial growth of γ-InSe and α, β, and γ-In$_2$Se$_3$ on ε-GaSe. *2D Mater.* 2018 **5**(3), 035026 (2018). https://doi.org/10.1088/2053-1583/aac479.

[24] C. Julien, M. Eddrief, K. Kambas, and M. Balkanski. Electrical and optical properties of In$_2$Se$_3$ thin films. *Thin Solid Films* 1986 **137**(1), 27–37 (1986). https://doi.org/10.1016/0040-6090(86)90191-4.

[25] J. O. Island, S. I. Blanter, M. Buscema, H. S. J. van der Zant, and A. Castellanos-Gomez. Gate Controlled Photocurrent Generation Mechanisms in High-Gain In$_2$Se$_3$ Phototransistors. *Nano Lett.* 2015 **15**(12), 7853–7858 (2015). https://doi.org/10.1021/acs.nanolett.5b02523.

[26] Raymond T. Tung. The physics and chemistry of the Schottky barrier height. *Applied Physics Reviews* 2014 **1**(1), 011304 (2014). https://doi.org/10.1063/1.4858400.
25